\title{Towards evaluating emergent behavior of the Internet of Things using large scale simulation techniques}
\author{
Stig Bosmans\textsuperscript{1}, Siegfried Mercelis\textsuperscript{1}, Peter Hellinckx\textsuperscript{1} and Joachim Denil\textsuperscript{2}\\ [15pt]
University of Antwerp\\
imec, IDLab\textsuperscript{1}\\
Flanders Make\textsuperscript{2}\\
Groenenborgerlaan 173 \\
Antwerp, Belgium \\
\{stig.bosmans, siegfried.mercelis,\\
peter.hellinckx, joachim.denil\}@uantwerpen.be\\
}
\date{\today}

\documentclass[12pt]{article}
\usepackage{graphicx}
\usepackage{wrapfig}
\begin{document}
\maketitle

\begin{abstract}
With the increase in Internet of Things devices and more decentralized architectures we see a new type of application gain importance, a type where local interactions between individual entities lead to a global emergent behavior, Emergent-based IoT (EBI) Systems. In this position paper we explore techniques to evaluate this emergent behavior in IoT applications. Because of the required scale and diversity this is not an easy task. Therefore, we  mainly focus on a distributed simulation approach and provide an overview of possible techniques that could optimize the overall simulation performance. Our focus is both on modeling and simulation technology. 
\end{abstract}

\section{Introduction}
\label{sec:intro}

In recent years, there has been a significant increase in the number of internet-connected devices. By 2020 it is expected that up to 50 billion devices will be connected to the internet. This Internet of Things (IoT) will create novel services that will highly impact our way of life in areas such as transportation, shopping, health and safety.  
With IoT becoming more mainstream and with the rise in the amount of devices getting interconnected, the complexity and scale of the IoT landscape will largely increase. The interoperability between IoT devices and actuators of all sorts will prove to be vital for future IoT applications. 
As a result of the increasing scale and diversity and because of new IoT architectures such as edge computing, the concept of emergent behavior in IoT systems is gaining more attention within the IoT community. Mataric et. al. \cite{mataric1993designing} defines emergent behavior as a collection of actions and patterns that result from local interactions between elements and their environment which have not been explicitly programmed. It is loosely based on the emergent behavior that is observed in bird flocks where birds apply three local rules which result in emergent flocking behavior (e.g. remain x distance to neighbor birds). D. Roca et. al. \cite{roca2016emergent} argues that this emergent behavior will lead to improved scalability, interoperability and cost efficiency of ultra large scale IoT systems as opposed to traditional approaches that heavily rely on extensive programming of explicit behaviors. This approach particularly benefits the IoT areas which require the interaction of an enormous amount of devices where relying solely on a centralized architecture is insufficient, such as smart power grids, autonomous car flocking and smart traffic lights. We will refer to this type of systems as emergent based IoT (EBI) systems in the remainder of this paper. Building such emergent-based IoT systems is complex as the system is composed from different heterogeneous and autonomous components. Examples of such components can range from actual IoT sensors sharing observations, towards smart actuators such as smart traffic lights that make decisions based on interactions with their counterparts and by observing sensor data. However, a human actor is also a component in an IoT system. From one perspective a human actor can be seen as a simple data generator, e.g. by walking around with a GPS-sensor embedded in their smartphone, they could broadcast location-data to an IoT middleware. From another perspective they can play a very active role in the IoT system by for example generating evolving traffic patterns which will influence the behavior of smart traffic lights.
All these various components and their individual behavior, their interactions and their goals need to be considered and evaluated when developing an EBI system. We believe that EBI systems will be the standard for future decentralized large scale IoT applications.
Because of the complexity, the large-scale and decentralized property of these EBI systems and their components, testing, validation and calibration will be vital in successfully developing such systems. In practice, IoT tests are either performed in small-scale/lab-based prototype setups or based on software testing approaches. The limited scaling capabilities of these testing approaches make it very hard to properly test EBI systems as a whole. EBI systems introduce a new component to test, being the global vs local behavior. Traditional testing techniques would not be able to test this global behavior at scale. Furthermore, the testing of the emergent behavior of IoT has been largely ignored in the literature. In this position paper, we investigate the challenges in testing EBI and look for solutions, methods and techniques mainly based on large-scale and distributed simulation based testing of IoT. The rest of this paper is organized as follows. Section 2 positions the challenges and needs in testing EBI systems, the next section provides an overview of possible techniques to improve performance and overall simulation scalability, in section 4 we further investigate state-of-the-art techniques that could be used in the scope of simulating EBI systems. Section 6 presents a discussion of the remaining challenges of real-time testing of IoT and how the presented techniques could help to solve these. Finally, in section 7 we conclude our work. 



\section{Challenges in evaluating emergent behavior}
\label{sec:challenges}

In the context of this paper, we are only interested in system-level evaluation of the EBI. This means that we are interested in observing the overall, global behavior posed by the application and we are unaware of the technical implementation details such as network and the hardware and software of the different sensor nodes.  

We distinguish three different scenarios where evaluating the emergence of behavior is necessary: 
\begin{itemize}
\item \textbf{Validation of emergent behavior}: We want to make sure that the local behavior and rules imposed by the application lead to the preferred emergent behavior. 
\item \textbf{Calibration of emergent behavior}: Calibration of the application needs to be done by tuning the various parameters in the IoT application, e.g. the value of an incentive. Therefore, an optimization of  emergent behavior has to be done by monitoring the impact the local rules and incentives have until the global, emergent behavior converges to a desired behavior. In other words, design-space exploration is needed to obtain optimized parameters for the simulation.
\item  \textbf{Integration testing and deployment}: Integrating the EBI applications in the real-world has a significant impact and therefore requires an incremental means to accept and deploy the application. Ideally, this translates to running the simulation models in parallel with the IoT middleware and the actual nodes operating in the real world. The number of nodes can be increased gradually to fully deploy the application in the field. For example, in a first integration test, a user could directly interact with a real prototype and with the simulated environment. Further tests increase the number of users in a local environment. 
\end{itemize}

The testing of EBI systems for each of these scenarios poses significant challenges because of the scale and diversity of these systems. We differentiate between three major methods that are relevant in this work:
\begin{enumerate}
\item \textbf{Real-world evaluation: }  This method relies on deploying the system in a real environment. Often the scale of such tests are limited to small lab-based setups. This has as an advantage, that the tests are easy to deploy. However in practice, this type of testing should only be used for testing the behavior of a single IoT device. Formulating reliable conclusions  about the validity of the behavior of the IoT system as a whole is unfeasible. 
Instead, there are a few large-scale, real-life IoT testbeds which allow tests to be executed at a more appropriate size. An example of such a testbed is the Smart Santander projects \cite{sanchez2014smartsantander}. It is a city-wide, real IoT testbed. It offers many thousands of IoT devices and interconnected gateways. It offers great opportunities to effectively test certain aspects of IoT devices deployed in a real environment. However, many EBI systems, such as the smart traffic light example, rely on custom devices and applications which would be too costly and too difficult to include in existing testbeds.

\item \textbf{Simulation-based evaluation:} A simulated environment consisting of a wide range of virtual IoT devices interacting with each-other and with the IoT middleware under test can be seen as a very effective testing solution. Although real testing is often desirable, simulation testing is a more flexible and cost-efficient approach. It allows for a more controlled and reusable environment that can be tweaked much easier. Within the IoT domain simulation testing is most often used to test technical aspects of the system such as network related features, power consumption etc. The most well-known examples of such simulators are NS-3 \cite{carneiro2010ns} and Omnet++ \cite{developer2010omnet++}. Various IoT operating systems such as Contiki and TinyOs also offer their dedicated simulator, Contiki \cite{dunkels2004contiki} and Tossim \cite{levis2003tossim} respectively. Some simulators are focused on testing more high-level IoT setups such as the iFogSim \cite{gupta2017ifogsim} which is used to test IoT edge or fog architectures. Another example is the CupCarbon simulator \cite{mehdi2014cupcarbon} which is used to simulate Smart City environments. Finally, D'Angelo et. al. demonstrate the use of the Gaia/Artis 
specifically to run large-scale IoT simulations \cite{d2016simulation}. However, most of the state-of-the-art IoT simulators do not have the necessary scalability requirements (more than 100.000 simulation entities) to simulate such large-scale IoT environments and to specifically evaluate emergent behavior. 
\item \textbf{Hybrid evaluation:} In order to facilitate the real-time testing approach explained above, a hybrid testing can be used which combines a small-scale real life environment with a virtual simulated environment. This way, actual IoT prototypes can interact with virtual devices, operating both as a single system able to generate emergent behavior. This could be particularly interesting for pilot projects or proof of concepts, where the scale is still too limited in order for emergent behavior to arise. Furthermore, it combines the advantages of both real-life testing and simulation testing. 
\end{enumerate}

Relying solely on real-world evaluation environments such as the Smart Santander project to test emergent behavior is costly and unpractical. Although these testbeds offer a number of ready-to-use sensors and IoT devices, deploying an emergent-based application on such an environment requires a significant amount of costly changes. Each EBI system will need a number of specialized sensors and actuators that need to be deployed, installed and linked to the available smart city infrastructure. Making such investments is impractical at the early stages of a project. Therefore, simulation or hybrid approach have a much smaller threshold to use. 

However, our observations detailed above show that the major challenge lies in the modeling of the global behavior and the scalability of the simulation framework. Furthermore, in all the different scenarios, there are different requirements on the accuracy of the simulation. On the one hand, validating the behavior of the EBI requires accurate results on the global behavior of the system. On the other hand, during calibration the parameter space of the application is explored with less accurate simulations to find good configurations of the application which are validated afterwards with detailed simulations.

\section{Overview}

The main challenge in simulating emergent behavior is related to performance. In order to obtain some level of emergent behavior a great amount of interacting entities is required. Running such a simulation on a single device wouldn't be possible, instead distributing to multiple servers is a better approach to cope with the necessary scale. Example implementations are described in the Parallel and Distributed Simulation (PADS) methodology \cite{fujimoto2000parallel} and in the IEEE 1516 standard. 
Facilitating the proper execution of these simulation entities in a simulated environment still requires several performance optimization techniques in the simulation kernel. Preferably, these optimizations can be performed transparent to the simulation modeler. However, this won't be possible all the time. In this sections we provide an overview of various optimization techniques that can be applied dynamically or statically. A dynamic optimization can be applied at run-time and can be context-dependent, while a static optimization needs to be applied up-front. We made a break-out of major performance optimization techniques in figure \ref{fig:performance} below. Note, that this list of techniques isn't complete and is the current result of on-going study.

\begin{figure}[h!]
\centering
\includegraphics[width=\textwidth]{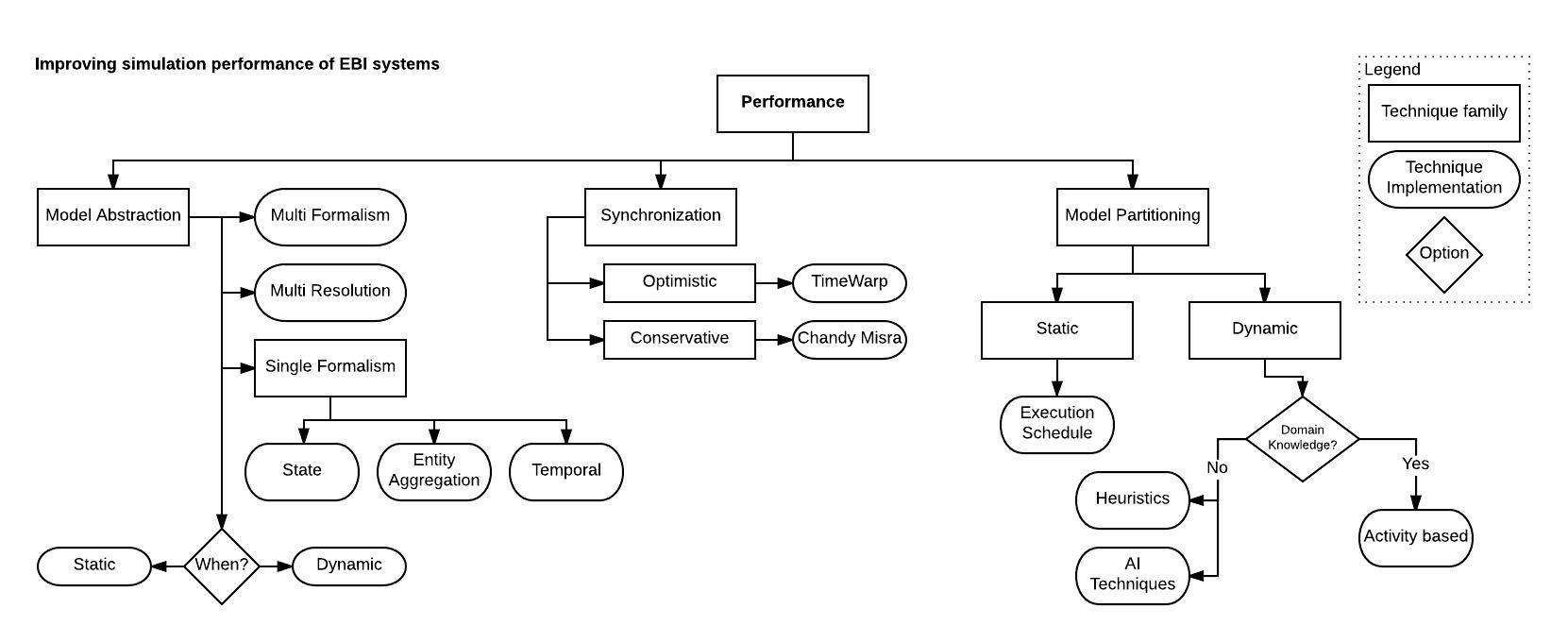}
\caption{Overview of simulation performance optimization techniques}
\label{fig:performance}
\end{figure}
\section{Possible Solutions when Modeling EBI}

An important aspect of generating emergent behavior is to properly describe the behavior of the individual elements of the real-world within the simulation model. Furthermore, once these systems are modeled we show which modeling techniques can help us in achieving a scalable EBI simulation framework.


\subsection{Model Representation}

In order to model emergent behavior the agent based modeling (ABM) paradigm seems the most appropriate ~\cite{macal2005tutorial}. With ABM, a bottom-up modeling approach is applied. Instead of modeling the global expected behavior, the modeler describes the behavior of the individuals. The ABM paradigm allows these individuals to interact. Eventually, these interactions will lead to a global behavior. As the EBI application are deployed in a similar matter, the simulation engineer has to model the different entities involved in the simulation. While this seems trivial, the individuals require detailed models on their decision making processes. 

Other formalisms can also be appropriate. Discrete-event formalisms such as DEVS~\cite{zeigler2000theory} model the behavior of a system using a timed sequence of "events" either as input to a system or as a timeout within the system. These events cause instantaneous changes to the state of the system. For example, when simulating traffic behavior, agent based modeling might not be the most appropriate formalism. Lots of research has been done in the domain of queuing theory which might be a better, less computationally expensive formalism. 
Atomic discrete event models can be combined together into a couple model. Furthermore, extensions exist that couple these models in a grid, e.g. Cell-DEVS~\cite{wainer2002cd++}. These types of cellular automata can have an advantage to model geographical areas. E.g. the city is divided into a grid of cells where each cell models a part of the city. 

Finally, differential equations can help to model rates of changes of properties within the IoT setting. E.g. traffic flows can be modeled with a differential equation. Different equations can also be coupled together using co-simulation techniques to allow for a divide-and-conquer approach to model complex systems. 

However, simulating such a large model of our IoT application is too complex. Model abstraction techniques should decrease the computational cost of individual models by “simplifying” the representation of such a model while maintaining the original behavior as accurate as possible. As a result, the overall computational complexity can be optimized. The main objective of applying model abstraction is to make a proper trade-off between computational complexity and accuracy. Extensive taxonomies of model abstraction techniques exist in literature~\cite{frantz1995taxonomy,caughlin1997summary}. In this work we will only discuss the most relevant techniques.
As demonstrated in figure~\ref{fig:performance}, we differentiate between two major categories, one based on multi-formalism model abstraction and another based on single-formalism abstraction. We than discuss how we can use these techniques in an IoT simulation environment. 

\subsection{Abstractions within a single formalism} 

These abstractions are applied on the model without changing the formalism to describe it. We define three types of abstractions that can happen within a single model. We will apply these types of abstractions on the individuals of the ABM but they can also be applied to models in other formalisms as well. 

\paragraph{State abstraction} 
This technique abstracts the state of a single individual. As the decision processes of individuals can be very complex, certain parts of the decision making process that an individual exhibits can be ignored without compromising the global behavior of the simulation, e.g. the individual also reasons about other properties that do not influence the global behavior, or higher-order reasoning with little to no impact on the decision of the agent.

The process of creating a more abstract individual from a detailed one can be done manually. This requires insight into both the decision making process of the individuals as into the application that generates the emergence of behavior. Therefore, more automatic techniques might be more appropriate. 
An example technique that can be used for this is metamodeling (as in surrogate modeling). Caughlin et. al. defines a metamodel as a projection of the original, high-fidelity model onto a subspace defined by new constraints or regions of interest \cite{caughlin1997summary}. In practice, a metamodel is a mathematical approximation of a complex model. The original model is treated as a blackbox, and the metamodel operates as a surrogate model that replaces the original. To come to such a metamodel a detailed analysis of the input/output mapping of the original model has to be performed. Based on this analysis a surrogate metamodel should learn to represent a similar mapping. Preferably, the surrogate model is more abstract and is less computationally complex. 
Various methods have been used for the development of metamodels such as polynomial regressions, radial basis functions (RBF) and others \cite{barton1994metamodeling}. Most of these techniques try to approximate input functions or data.
Since the goal of metamodeling is to map an input value to a specific output value we believe other supervised learning techniques could be used as well. For example, P. Symonds et. al. demonstrate that neural networks (NN) could perform up to 15\% better compared to classic RBF approaches \cite{symonds2015performance} \cite{gore2017augmenting}. We believe that evaluating various state-of-the-art deep neural networks could lead to even better results. We consider this as a promising area for further research. 

\paragraph{Entity Aggregation}
Another option to reduce the computational complexity of simulation entities is Entity Aggregation. The idea of entity aggregation is to combine multiple low-level simulation entities with a single high-level entity while preserving the collective behavior. For example, when simulating emergent behavior in a smart power grid application, a single agent could represent a single household but depending on the required level of accuracy a single agent could also represent an entire neighborhood. 
Rodriguez et. al. demonstrates that NN's could be used to transform a collection of low-level entities to an aggregated model \cite{rodriguez2008building} . From an architectural perspective, a mechanism in the simulation kernel can be created to detect and analyze clusters of agents that interact a lot and pose homogeneous behavior. Using metamodeling techniques such a cluster of agents can then be replaced by a single surrogate model. Conversely, the modeler can also manually model an aggregation of individual simulation agents.

\paragraph{Temporal Abstraction}
Finally, temporal abstraction can be used to limit the granularity of simulation entity state updates over time. Since, each state update results in a cascade of simulation events the overall computational cost could be significantly reduced. 

Engineers can themselves easily change the temporal granularity (time-step) of their simulation. However, choosing a too large time-step can result in an unstable simulation.  Automatic techniques to adapt the time-step also exist. E.g. for solving differential equations, several techniques are available that automatically adapts the time-step of the simulation based on the estimated error, e.g. Runge-Kutta 4-5 embeds a higher-order method in its solver to estimate the error and adapt the time-step. 

\subsection{Multi-formalism Modeling}
Changing formalisms during abstraction can help to achieve better performance results as we can leverage the strengths of different types of formalisms. For example, as explained in section 4.1, in the domain of traffic simulation, queuing theory might be a more appropriate, less computationally expensive formalism compared to agent based modeling. The formalism change allows for an easier state and aggregation abstraction.
The disadvantage of applying multi-model abstraction is that this can't be automated, the modeler will be responsible to model the same behavior in different formalisms. 

\subsection{Multi-resolution Modeling}
Multi-resolution modeling can be used to vary between model abstraction levels (and possibly formalisms) within the same simulation. The advantage of this is that we can increase abstraction levels in simulation areas where less accuracy is required. This will consequently reduce the computational complexity of the simulation. Oppositely we could also increase levels of detail in certain simulation areas when an increased level of accuracy is required. The multi-resolution model can be static or dynamic.

When applying a static approach, the most appropriate formalisms and levels of detail have to be specified upfront and will be used during the entire course of the simulation. Due to the dynamic nature of IoT applications, the potential computational benefit is limited.
Conversely, with a dynamic approach we can switch between various levels of abstraction during the course of the simulation. This can be done by predefining when and where an abstraction switch should occur. Domain knowledge is necessary to detect and abstract/refine the (part of the) model.

Learning approaches are also possible to switch levels of abstraction, by learning areas of opportunity and applying the necessary level of abstraction. For this, an additional mechanisms in the simulation kernel are required that detects areas of opportunity based on various parameters. For example, simulation entities that have a limited amount of interaction with the rest of the simulation could be marked as candidates for an abstraction level increase. 
The advantage of such an approach is that we can adaptively detect opportunities to switch between abstraction levels in order to decrease necessary computational resources or oppositely, use all available computational resources.

Applying a dynamic multi-resolution modeling approach leads to additional challenges that need to be taken into account~\cite{multires2015}. For example, reinitializing a more abstract model from  the current state of the original model is possible as there is enough information available in the original. However, when switching from a more abstract to a refined model, extra information is required. This should be mend by modeling additional domain-knowledge to fill this information gap.

\section{Possible Solutions when Simulating EBI}
In this section we will discuss techniques that could improve the performance capabilities of the simulator itself. Most of these techniques have been studied in various domains such as parallel and distributed simulation (PADS) and distributed discrete event simulation.

\subsection{Simulation Architecture}
Most of the current IoT simulators have an underlying monolithic architectures which limits its scalability capabilities. D'Angelo et. al. recognizes this limitation in state-of-the-art IoT simulation in their work. Instead, they propose their custom built large-scale simulator Gaia/Artis which is based on the parallel and distributed computing (PADS) paradigm. PADS allows a simulation to be executed among multiple distributed devices or Physical Execution Units (PEU’s) , which will significantly improve scaling capabilities.
Each PEU consists of a collection of logical processes (LP). An LP represents a part of the simulation and
contains a collection of simulation entities (SE). A simulation entity, as the name implies, represents an in-
dividual simulation model or agent (in the context of agent based simulation).
A disadvantage of PADS and distributed simulation is that it leads to additional difficulties with regards to synchronization, simulation partitioning and transparency. Extensive research has already been done to cope with these challenges. Although the PADS methodology to our knowledge hasn't been applied to evaluate real-time, simulation-based IoT systems, many of its challenges are comparable. 

\subsection{Model Partitioning}

Model partitioning aims to optimally distribute agents across multiple servers to decrease the overall communication cost. A significant part of the computational interaction cost is characterized by remote communication across multiple PEU's \cite{bononi2006proximity}. This computational cost of remote communication is much higher than the local communication cost between SE's  located in the same region. Much efficiency can be gained by analyzing communication patterns and optimizing the distribution of the simulation entities, so that local communication is maximized and as a result the communication cost will be lower. 


\paragraph{Static Model Partitioning:}
We make a distinction between a static and a dynamic approach. With a static partitioning approach it is assumed that the interaction dynamics between SEs remain unchanged, as soon as these dynamics change over the course of the simulation the benefit of the partitioning decreases. For example, a fixed partitioning could be configured that specifies the migration times of individual simulation models. Such a schedule could be performing well at first, but as soon as the communication dynamics change it will lead to suboptimal simulator performance \cite{van2014activity}. Instead a fixed execution schedule that specifies exact migration times or preferred clustering schemes of individual simulation entities. 
However, in the domain of IoT we can assume the presence of stochastic dynamic nodes and devices. This will inherently lead to unpredictable changes in interaction dynamics. Furthermore, because of the required scale to simulate emergent behavior we can conclude that static partitioning methods are impractical and not preferable in order to simulate EBI applications \cite{d2011parallel}. 

\paragraph{Dynamic Model Partitioning using Heuristics:}
When compared to static partitioning, the dynamic partitioning methods will be more adequate for IoT use cases, as they are able to automatically migrate simulation entities when their communication patterns evolve. 
In the context of PADS, many research has been done in the area of dynamic simulation partitioning. Most of these techniques rely on generic solutions based on heuristics. In most cases these heuristics don't have a complete view on the global simulation state because this would be too computationally expensive. Instead most heuristics rely on limited data directly available on the individual hosts. 
D'Angelo proposes three heuristics \cite{d2017simulation}, each one is slightly different and can be used for different types of simulation use cases. 
\begin{itemize}
\item Heuristic 1: The first heuristic relies on a sliding window approach of the last k timesteps. For each simulation entity the ratio between external interactions versus internal interactions is evaluated. When the ratio exceeds a predefined threshold the SE can be migrated to the LP it communicated with the most when at least a minimum amount of timesteps have passed since the last migration of the SE. 
\item Heuristic 2: This heuristic is very similar to the latter, but in this case the sliding window is based on the last k interactions instead of the last k timesteps. As a result SE's that have less interactions will be marked sooner as migration candidates. This is because old interaction events will still impact the migration decision, opposed to the timestep windowing approach where these would possibly have been discarded long ago. 
\item Heuristic 3: The third heuristic is very similar to the second, but instead of evaluating the interactions each timestep, which is very computationally intensive, the evaluation will only occur when at least a certain amount of interactions occurred since the last evaluation. 
\end{itemize} 

The heuristics discussed above prove to have very positive impact on the overall performance and the scalability of the simulations.
In other work, various algorithms are proposed that focus more on the distribution of simulation workload instead of communication. For example, Boukerche et. al. propose a solution to dynamically distribute the SE's in the context of a distributed simulation that leverages the Chandy-Misra null messages conservative synchronization algorithm \cite{chandy1979distributed} \cite{boukerche1997dynamic}. As a result, a significant decrease in simulation runtime is achieved.  

\paragraph{Dynamic Model Partioning using Domain Knowledge:}
A possible disadvantage in the heuristics approach is that it doesn't take domain knowledge into account. This can lead to a number of undesired side effects. For example, migrations might occur that need to be undone in a later phase. This could result in a higher computational cost (introduced by these consecutive migrations) than the obtained gain which was only temporal. The heuristics presented above do try to prevent oscillating migrations by introducing a threshold that prevents immediate undoing a migration. However,  another solution would be to inject domain knowledge that could prevent these unwanted migrations to occur in the first place. Furthermore, it could perform more optimal migrations that couldn't have been achieved by solely analyzing local communication patterns of individual SEs.

Van Tendeloo and Vangheluwe demonstrate in their work that significant performance improvements can be obtained by injecting domain knowledge in their Python based distributed DEVS simulator PythonPDEVS \cite{van2014activity}. 
Their work builds further on the abstract notion of activity introduced by Muzy et. al. which represents a measure for a number of events in the context of a Discrete Event Simulation system \cite{muzy2010activity}. This activity concept can refer to various resources such as time, memory or energy which are relevant in a DES system. In the context of performance optimization the time resource will be most relevant. One can look at activity from various perspectives, either from a computational load perspective or from a communicational perspective. The heuristics discussed in the previous paragraph looks at activity from a communicational perspective whereas activity-based on the computational perspective.
The concept of activity prediction allows simulation models to provide hints to the simulator kernel about both their current and anticipated activity and how they should be distributed. Consequently, these hints will be exploited to decide when a SE migration should occur. This approach leads to a more optimal distribution of computational load across servers. The idea of activity prediction based on domain knowledge is also relevant when looking at the activity concept from a communicational load perspective, in that case, the goal would be to reduce the overall communicational load over the network.
A disadvantage of this approach is that a level of transparency has to be sacrificed, breaking down the boundary between model and simulation engine in order to obtain better performance. 


\subsection{Synchronization}

Finally, another important decision that needs to be considered when developing an EBI simulator is synchronization. This section gives an overview of the most relevant techniques described in state-of-the-art research.

Synchronization is an important topic in the field of distributed simulation. Its goal is to maintain validity of a distributed simulation system by preventing causal inconsistencies. Such an inconsistency occurs when events that depend on each-other are executed in the wrong order \cite{lamport1978time}. For example, imagine an event B that depends on the results of an event A, if event B were to be processed before the execution of event A, a causal inconsistency occurs and simulation results could become invalid. 
Various synchronization techniques are discussed in literature to maintain this causal consistency, a distinction is made between two major categories 1) optimistic synchronization techniques and 2) conservative optimization techniques. Each technique will lead to different performance results. 

\paragraph{Conservative Synchronization}
One way to maintain causal consistency among various LP's in a distributed simulation is to prevent a simulator to move to the next event only when it's sure that no LP's will insert an earlier event or a next event to be removed \cite{nicol1996principles}. This approach is called a conservative synchronization approach.
The best known conservative techniques are based on the Chandy Misra Bryant algorithm \cite{misra1986distributed}. In their work each LP should have an individual communication channel for all other LP's it communicates with. Each LP is assumed to post timestamped events in the right order. An LP is only able to accept and process events with the lowest time-stamp as soon as all messages in all channels are received.
This could lead to deadlocks when some LPs are not generating events, leading to some empty channels and an LP consequently remaining in a waiting/blocking state. As a solution, the idea of null messages is added, where null messages can be send to channels when no events are assumed for a particular LP. 
The CMB technique allows the simulation to progress at various rates, however it introduces high overhead because of the null messages  in terms of computational overhead and network load \cite{d2014new}.

\paragraph{Optimistic Synchronization}
The conservative synchronization method blocks the simulator and results in some LPs waiting in an idle state on other, slower LPs, which is not always necessary. Instead, the optimistic synchronization method introduces a non-blocking approach where each LP continues on its own rate, not waiting for other LPs until an inconsistency occurs, e.g. an event is received with a time-stamp earlier than the local time of the LP. When such a inconsistency is found the simulator will turn back its state to a checkpoint prior to the time-stamp of the received event before continuing processing. As a result, the simulation remains valid. 
This non-blocking optimistic synchronization mechanism as described above, is called the time warp protocol, first introduced in Jefferon's paper \cite{jefferson1985virtual}. In many cases it could lead to better simulation performance and a reduced execution time.

\section{Discussion}
Most of the performance optimizing techniques discussed in the above sections are focused on a full simulation-based testing approach of EBI systems. However, our ambition is to eventually enable a hybrid simulation based testing environment. The high-level architecture of such a system is depicted in figure \ref{fig:sim-based-deployment} below. A large-scale simulated environment would be combined with an actual environment containing a prototype setup limited in scale. Both environments are able to communicate with a supporting middleware system. This way a full-scale EBI system can be deployed, tested, calibrated and validated as a whole without the need to do a full deployment in an actual environment. However, this will have a number of consequences to the simulation technology being used. A distributed simulation environment relying on a PADS-like architecture would be preferable. However, the methods that can be used are limited. An optimistic synchronization approach is not realistic because we could not allow a simulation model to rewind its state after an inconsistency is found as this would break the required real-time property. As a result, only conservative synchronization methods are feasible. Furthermore, we need to have a guarantee that all SE states are updated in due time while maintaining consistencies, the blocking nature of the conservative synchronization mechanism might result in state updates not being processed at the proper rate. 

\begin{wrapfigure}{R}{9cm}
\includegraphics[width=.55\textwidth]{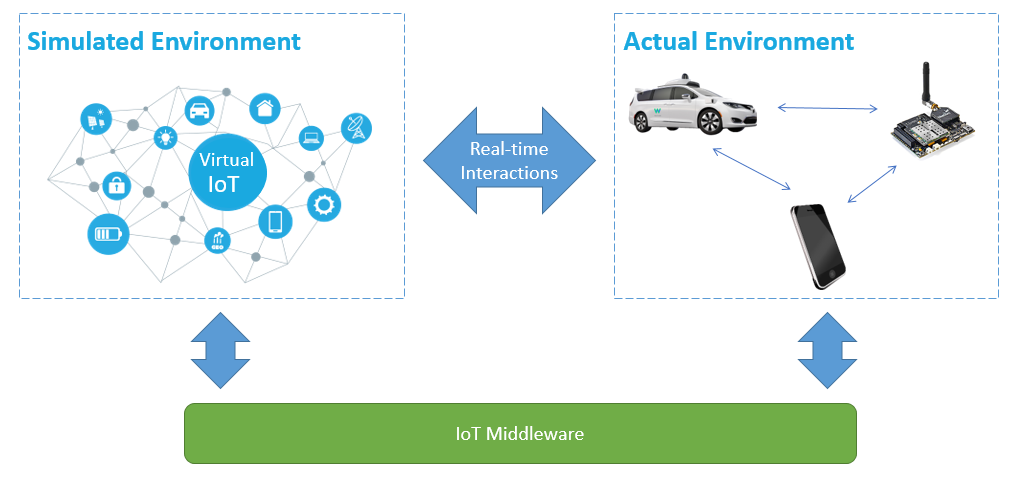}
\caption{Hybrid evaluation scenario}
\label{fig:sim-based-deployment}
\end{wrapfigure}

Additionally, in the hybrid evaluation scenario a state synchronization is required in the context of EBI applications. This mechanism needs to ensure that actual prototype in the real-world perceives similar data as the virtual, simulated devices. Therefore, the state of the simulation needs to be synchronized with the state of the real-world For example, in the case of an EBI systems related to air pollution, measurements of air pollution in the virtual environment need to match those measured in the real world. 

Finally, from a model partitioning perspective, we must make sure that SE migrations between LP's maintain the availability and responsiveness of the SE, e.g. it could not be allowed that a SE remains unresponsive during the process of migrating. We identified many existing techniques that could help to improve the performance and overall scalability of large-scale EBI IoT systems. In further research we will focus on adding IoT and EBI related domain knowledge to the aforementioned techniques and to better understand in which cases which techniques are most appropriate. 
\section{Conclusion}
In this position paper, we identified the challenges that arise when trying to evaluate emergent behavior in Internet of Tings applications. A type of application, where decentralized behavior is key.  Furthermore, we proposed the idea of a hybrid simulation environment that combines real-world and simulated nodes for testing and deployment of these emergent behavior IoT systems. 
As a solution, we propose a distributed simulation-based approach.
However, relying solely on a simulation architecture  is not enough to cope with the scalability challenges of simulating emergent behavior. Instead, we present an overview of state-of-the-art techniques that could help to improve simulation performance and overall scaling capabilities.  In further research we model and leverage the IoT domain knowledge on the techniques mentioned in this study to develop a methodology for evaluating emergent behavior in IoT systems. 

\bibliographystyle{abbrv}
\bibliography{ms}

\end{document}